**Title**

# Thermal threshold for localized Blood-Brain-Barrier disruption

**Authors**

Sébastien Bär[1,2,3†*], Oliver Buchholz[1,2†], Christian Münkel[1,2], Paul Schlett[1], Pierre LeVan[4], Dominik von Elverfeldt[2,3], Ulrich G. Hofmann[1,2]

**Affiliations**

[1]Section for Neuroelectronic Systems, Department of Neurosurgery, University Medical Center Freiburg, Freiburg, Germany

[2]Faculty of Medicine, University of Freiburg, Freiburg, Germany

[3]Division of Medical Physics, Department of Diagnostic and Interventional Radiology, University Medical Center Freiburg, Freiburg, Germany

[4]Department of Radiology and Hotchkiss Brain Institute, Cumming School of Medicine, University of Calgary, Calgary, AB, Canada

[*]sebastien.baer@uniklinik-freiburg.de
[†] These authors contributed equally to this work

**Abstract**

The Blood-Brain Barrier is the gatekeeper of the CNS. It effectively shields the brain from blood-borne harm but simultaneously represents a significant challenge for treating neurological diseases. Altering its permeability enables increasing the local drug concentration and thereby improving the therapeutic effect. Although permeability increase is achieved by raised tissue temperature, the determination of the thermal dosage suffers from imprecise thermometry during hyperthermia application. Knowledge of the thermal dosage is crucial for improving hyperthermia related interventions of the CNS. Here we show an approach to determine the thermal threshold for localized Blood-Brain Barrier disruption estimated by MR thermometry.

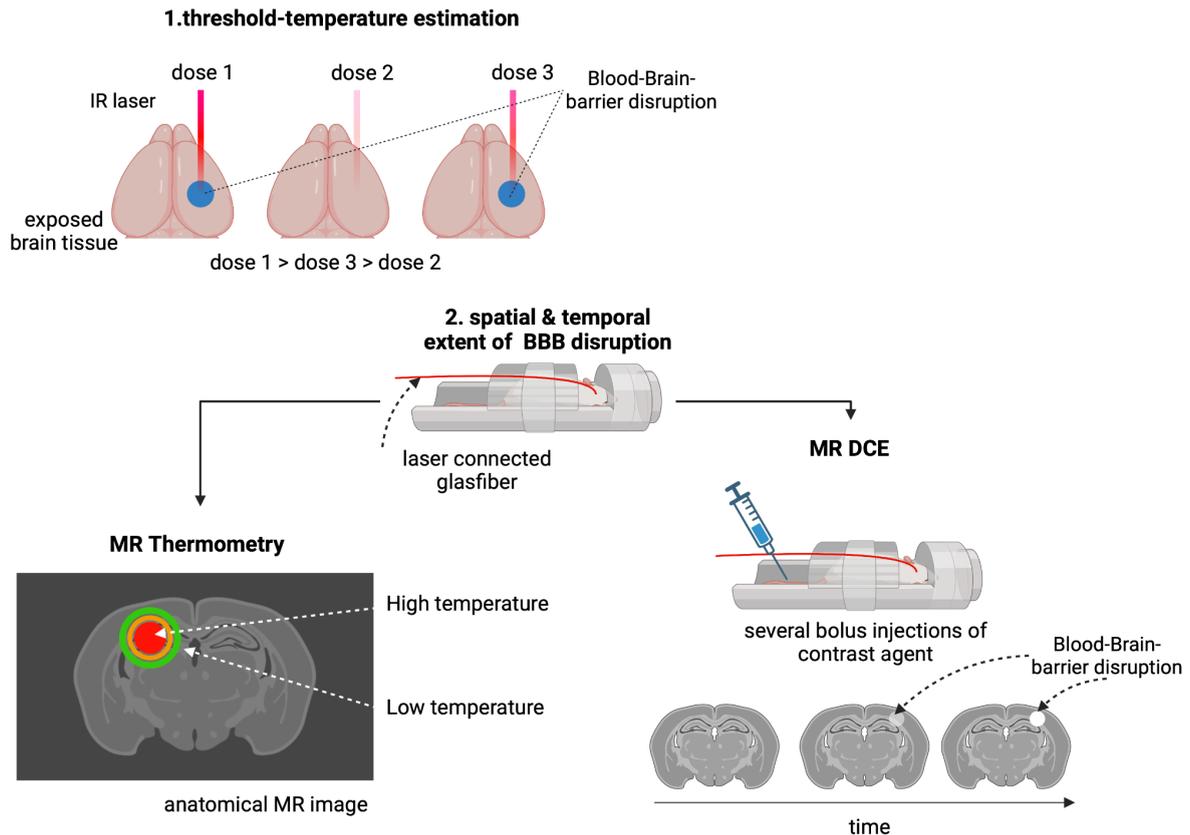

*Graphical abstract: Initial experiments were conducted to estimate the thermal dosage threshold for Blood-Brain-Barrier disruption. Different local tissue temperatures were achieved by IR laser illumination with different amounts of energy deposition. Increased Blood-Brain-Barrier permeability was determined by local dye extravasation at the illuminated tissue area. The spatial and temporal extent of Blood-Brain-Barrier disruption was subsequently quantified by MRI. IR laser powered glass fibers were implanted into the brain and the necessary thermal dosage for Blood-Brain-Barrier disruption applied. The thermal distribution during IR illumination was monitored by MR thermometry. In an additional experiment, the same thermal dosage was applied during multiple bolus MR DCE to monitor Blood-Brain-Barrier disruption following IR illumination. Relating results of both measurements yields temperature sensitivity maps of Blood-Brain-Barrier disruption for several points in time. (sketch designed with biorender)*

**Methods**: Using an IR laser ($\lambda$ = 1470nm) we showed that highly localized Blood-Brain Barrier opening can be achieved with mild to moderate hyperthermia. Non-invasive MR thermometry has been used to determine the temperature at the heating site. Blood-Brain Barrier opening has been monitored by DCE-MRI *in vivo* and post mortem via Evan's Blue extravasation.

**Results:** The Blood-Brain-Barrier permeability can be altered locally with minimal thermal dosages. Thus mild hyperthermia represents a promising approach to making the brain accessible for therapeutic interventions.

# Introduction

*The Blood- Brain- Barrier*

The complex mechanisms required for a functioning brain rely on maintaining delicate membrane processes in nerve cells, well balanced glia cell activity and intracellular homeostasis [1]. Unfortunately, the most complex organ of our body is not equipped to hold extensive energy, oxygen or nutrition storage [2]. Consequently, the required supply for various metabolic processes, most predominantly in the form of glucose and oxygen, as well as other crucial substances, must be transported on demand to the different areas of the brain [3].

This necessitates ubiquitous, far reaching and immediate cerebral blood supply, with the transport of any substance into brain parenchyma being tightly controlled and regulated [4]. Different tissue-circulation-interfaces separate the nervous system from the rest of the body, featuring distinct characteristics tailored to the metabolic and functional demands of the corresponding system [5]. Breakdown of either of these barriers has been implicated in several different pathologies and CNS-related diseases. The barrier system between the circulating blood and the CNS, the Blood-Brain Barrier (BBB), constitutes a bottle neck for treatments of neurological diseases as it tightly controls the entry of blood-borne substances into the brain [6,7]. This represents an important safety mechanism against external harm and regulates brain function but also prevents the vast majority of drugs from being transported to target regions [4].

Modulating the BBB to increase local drug concentrations has therefore been the focus of intensive research. Various approaches to facilitate passive or active transport through the BBB have been discussed in the literature including osmotic disruption [8–11], pharmacological modulation [10,12] or nanoparticle-mediated substance delivery [13–15]. Advances of these techniques suffer from adverse tissue expansion, insufficient selectivity and inadequate biocompatibility respectively.

One of the most promising advances uses focused ultrasound (FUS) to induce mechanical disruption at the BBB [16]. Several ultrasonic beams can be applied through the skull to deliver acoustic energy several centimeters deep into the brain [17,18]. FUS can be expanded by injecting microbubbles which consist of a lipid or albumin shell encapsulating a gas [19–21]. These microbubbles expand and contract rapidly when they enter the pressure field of the FUS application which creates mechanical forces at the site of the BBB and thereby disrupts it. Because of their size (ø 1-10 µm) [22] the microbubbles are confined to the vasculature enabling localized BBB alteration. FUS has been shown to increase the concentration of cytostatics [23,24] as well as therapeutic agents for neurodegenerative diseases [25,26] at target sites in the CNS. Recently, MR guided FUS has been used to increase BBB permeability in Alzheimer patients and has been shown to reduce the amyloid plaques associated with the disease in humans in clinical trials [27].

However, off-target effects, such as tissue damage or inflammation can be caused by cavitation and acoustic streaming due to the high-intensity waves used during FUS [28]. Individual variations in the anatomy and physiology of the brain can also influence the occurrence of off-target effects. Disparities in tissue density, skull thickness, and other factors can affect the propagation of ultrasound waves and lead to deviations in the distribution of acoustic energy [29]. The precise control and monitoring of ultrasound parameters are crucial to minimize off-target effects, and there is still a need for further research and optimization of the technique to ensure its safety and efficacy.

Extensive research is being conducted to explore additional approaches for disrupting the BBB that minimize the issues connected to the existing techniques.

The state of BBB permeability is, among other factors, dependent on the brain or body temperature [30]. Indeed, there is evidence that local heating can reversibly [31] open the BBB. To date however, there are two central challenges regarding hyperthermia applications for BBB disruption. First, the generalized application of heat either to the whole body or large portions of the brain leads to adverse off-target effects [32,33] comparable to the above described effects during FUS application. Secondly, monitoring of the temperature progression during hyperthermia application is challenging and therefore specific thermal dosage thresholds for BBB opening have not been convincingly described in the literature.

In the following study, IR laser illumination was applied directly to the brain parenchyma to investigate localized BBB disruption as a consequence of localized hyperthermia. Simultaneously, the temperature profile was retraced using MR thermometry enabling determination of a thermal dosage necessary for significant increase in BBB permeability. At this point, especially the investigation of the spatial and temporal characterization of temperature changes necessary for BBB opening are relevant.

*MR Thermometry*

With respect to non-invasive temperature monitoring the proton resonance frequency (PRF) shift method is one of the most commonly used mapping procedures for *in vivo* applications due to its local linear temperature dependency and its (quasi) independence from tissue type [34,35]. The PRF method is based on repeated phase map acquisitions using a standard gradient echo MRI sequence (FLASH). Local chemical shift effects introduced by temperature changes are detected as phase changes in the images. If the temperature corresponding to the first acquired phase map is known, the temperature changes over time can be obtained through equation 1:

$$\Delta T = \frac{\varphi(t) - \varphi(t_0)}{\gamma \alpha B_0 TE} \qquad \text{equation 1}$$

with $B_0$ the external magnetic field, $\gamma$ the $^1H$ gyromagnetic ratio, $\alpha$ the thermal coefficient and TE the echo time. $\varphi(t)$ is the measured phase in a voxel at a given time t and $\varphi(t_0)$ is the corresponding initially acquired reference phase at a known temperature. Since fat saturation pulses are used during the acquisition, only aqueous tissue signals remain in the phase map. In this case, $\alpha$ = -0.01 ppm/K can be reasonably assumed for all voxels in the image providing the reason for its tissue independence [36].

Thus, precise mapping of temperature changes due to IR heat deposition in a targeted area is possible with high spatial and temporal resolution.

*Multiple-bolus DCE*

Standard techniques for noninvasive quantification of BBB permeability rely on MRI measurements after injection of a gadolinium-containing contrast agent (CA) that does not pass through the BBB. Gadolinium (Gd) shortens $T_1$ and $T_2$ relaxation times, resulting in signal changes on MR images. A quantitative determination of the BBB permeability can then be performed using the so-called dynamic contrast enhanced (DCE)-MRI technique. In this technique, a time series of $T_1$-weighted images is acquired after injection of a bolus of CA. The rate of CA formation in brain tissue can then be inferred from a pharmacokinetic model comparing the $T_1$ signal in each voxel with the arterial signal. The entire time series thus yields a single value of BBB permeability. Therefore, to observe dynamic fluctuations of BBB permeability, further CA injections and acquisitions of $T_1$-weighted images are necessary. Also, as the terminal half-life of Gd-CA's in rats typically is around 15-20 minutes [37], these further DCE-MRI acquisitions need to take residual CA from the previous scans into account, which can be accomplished using calibration $T_1$ maps acquired before each DCE-MRI acquisition [38].

## Methods

*Cortical BBB opening by free IR laser illumination*

In order to re-examine literature values and determine a temperature range and exposure time sufficient for visibly increasing BBB permeability, several preliminary experiments were conducted prior to MR thermometry. The corresponding experimental setup has been described previously [39] and utilized a laser beam (1470 nm, 12 W, DILAS, Mainz, Germany) collimated by an optical lens assembly mounted on a 3-axis micromanipulator for beam positioning [40]. An experimental paradigm was established that reproducibly achieved stable and controllable heating of the brain tissue. The amount of pulses per second (frequency), the duration of the pulse (pulse width) and the maximal power per pulse (peak pulse power) determine the maximal temperature. Different pulse widths as well as different peak powers have been applied to achieve different maximal tissue temperatures. The IR illumination parameters for each sample point are depicted in the supplementary table S1.

The local brain tissue temperature was monitored using a high speed thermal camera (FLIR, A8303sc, custom macro lens).

IR-illumination with a spot size of 538 µm diameter was applied to 9 Sprague-Dawley rats (290-310 g) according to the permit G17/80 of the responsible Regierungspräsidium Freiburg and our clinic's Internal Review Board. Anesthesia was induced with Isoflurane at 4% in $O_2$ and maintained at 1.5 - 2% via a nose cone while the animals were mounted in a stereotactic frame (Kopf, Model 900LS, USA). The rat's heads were shaved and an incision was made to expose the skull. Four burr holes with a diameter of 1 mm were carefully and intermittently drilled above each hemisphere while cooling with saline[41] enabling the application of several thermal dosages directly on the brain. Locations were fixed at 2.5 mm lateral and 2.5 mm posterior to Bregma for the anterior ROIs and 5 mm lateral and 5 mm posterior to Bregma for the posterior ROIs [42] (see figure 1). Subsequent to IR illumination, EB was administered via the tail vein (4% EB in 0.9% NaCl, 2 ml/kg). Immediate blue coloration of the eyes, limbs and ears indicated successful administration. IR illumination induced EB extravasation was verified by applying no or very low energy to a random location per animal.

*Glass fiber implantation surgery and preparation for IR-Laser illumination during MR measurements*

Animal surgery was done similarly for both the MR thermometry and the DCE-MRI measurements. For each method, one Sprague Dawley rat was anesthetized analogously to the above described procedure. Subsequently, two burr holes were drilled (2.5 mm lateral and 5 mm posterior to bregma). The burr holes were then covered with a 3D printed guiding platform glued to the skull through which bare glass fibers were introduced ($\emptyset_{core}$ = 550 µm), FG550LEC, Thorlabs, USA). One of the glass fibers was connected to an IR laser (ML7700, Modulight Inc.,Tampere, Finland) while the other one served as control. This setup ensured similar positioning of the 2 glass fibers during measurements including location of the glass fiber termination and the angle of implantation (20º).

For the respective MR measurements, the animals were transitioned into an MR animal cassette (Bruker) while anesthesia was maintained through a nose cone.

Animal core temperature was monitored using a rectal thermometer and maintained at approximately 37 ºC for the duration of the experiments.

Following multi-bolus DCE-MRI measurements, EB was injected via the tail vein. After verification of sufficient EB circulation by discoloration of limbs, ears and eyes, the animal was sacrificed by isoflurane overdose and decapitation. The brain was extracted for subsequent verification of EB extravasation.

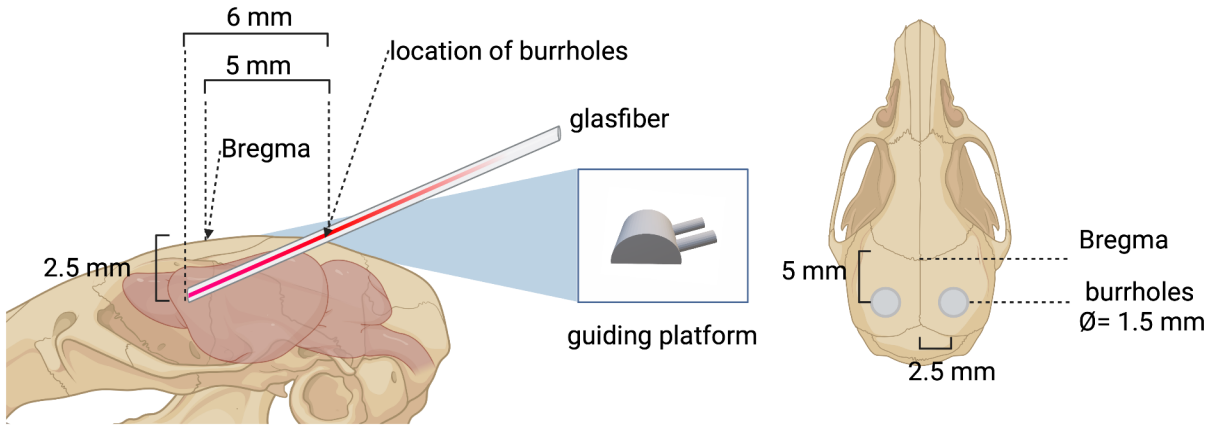

*Figure 1: Experimental set up during in vivo MR measurements: Two glass fibers were implanted through burr holes (ø1.5mm) via a 3D printed guiding platform. One glass fiber was connected to an IR laser to deliver heat at the target site, the other served as control. (sketch designed with biorender)*

*MR Hardware/Instrumentation*

All MRI measurements were performed at a 9.4 T, small animal scanner (Bruker Biospec 94/20, Bruker BioSpin, Ettlingen, Germany). For the phantom measurements a Bruker transmit-receive linear rat coil was used while for the in-vivo measurements a supplemental 4-channel rat head surface array coil was used for detection. A standard Gd-based MRI CA (Gadoteridol, ProHance, Bracco Diagnostics, Inc., Princeton, NJ) with a relaxivity of $r_1$ = 2.6 mM$^{-1}$s$^{-1}$ at 9.4T [43] was used for all in-vivo DCE experiments.

*Phantom MR-Thermometry*

In order to calibrate heating, we placed the glass-fiber end in a 50 ml falcon tube filled with 0.6% agarose. The concentration of 0.6% agarose was chosen for its similar viscosity to brain tissue [44]. Heat maps were acquired for different energy levels of the infrared laser beam. The laser was used in a pulsed-mode, enabling it to reach an almost constant temperature after only a few pulses, referred to as temperature plateau in this publication. In order to limit and control the mild temperature increase as much as possible, the lowest laser peak power as well as the shortest possible pulse width available were selected ($P_{peak}$ = 500 mW and τ = 1 µs respectively), inducing a pulse energy of 0.5 µJ. In order to obtain different temperature plateaus, we varied the duty cycle (DC) of the laser. DC in % is defined by the fraction of time during which the laser is actually emitting light (so the pulse width τ) per period Δt. As τ was chosen fixed, only Δt and with it the average power ($P_{av} = P_{peak} * τ/Δt$) was herewith varied.

As DC in % was the parameter presented on the laser's display, we decided to use both DC and $P_{av}$ for the characterization of the heating effects in this publication.

For the phantom experiments, DC was varied from 1.25%, 2.5%, 5%, 7.5% and 10%, corresponding to $P_{avg}$ of 6.25 mW, 12.5 mW, 25 mW, 37.5 mW and 50 mW respectively. The laser was turned on for 128 s. MRI heat maps consisted of phase maps acquired with a fat saturated 2D-RF-spoiled FLASH sequence. Parameters were: Repetition time (TR) = 80 ms, echo time (TE) = 10 ms, Flip angle (FA) = 15°, field of view (FOV) = 32 mm x 32 mm and 5 juxtaposed 500 µm-slices with a 250 µm or 500 µm in-plane resolution.

As the precision of ΔT is inversely proportional to TE (see equation 1), a relatively long TE (10 ms) is necessary to achieve a higher precision (~0.5 °C). At the same time SNR is exponentially decreasing with TE, limiting the spatial resolution (250 µm). Finally, as this spatial resolution limits the associated temporal resolution (10.24 s per phase map), we also acquired phase maps with 500 µm and 5.12 s spatio-temporal resolutions in order to investigate the heating behavior over time in the rat brain (plateau formation).

The number of acquired phase maps was adapted depending on the experiment length. The laser beam was started after at least 10 repetitions to ensure a constant temperature. The mean value of those first points was used to define the initial phase map $φ(t_0)$ corresponding to the reference temperature simultaneously measured by a temperature probe. In a post-processing step, the acquired phase maps were corrected for phase wraps and heating maps were calculated using equation 1.

In order to validate the resulting temperature maps, similar temperature profiles were measured with a thermal camera. For that, the laser connected glass fiber was guided through a platform ending in a depot of agarose (0.6%). The temperature of the agarose surface was subsequently measured with a thermal camera positioned perpendicular to the agarose surface (see supplementary figure S2). As the thermal camera can only measure 2D temperature maps from any structure's surface, 7 predefined locations of the glass fiber end were used to obtain heat maps at 7 different distances from the laser beam. In the thermal camera images, 250 µm ROIs were drawn at the glass fiber end site for comparison.

These dummy measurements were compared to post-mortem brain MR thermometry with DC values of 1% to 6% in 1% steps corresponding to a $P_{av}$ of 5 mW to 30 mW in 5 mW-steps were used.

*Multiple-bolus DCE-MRI*

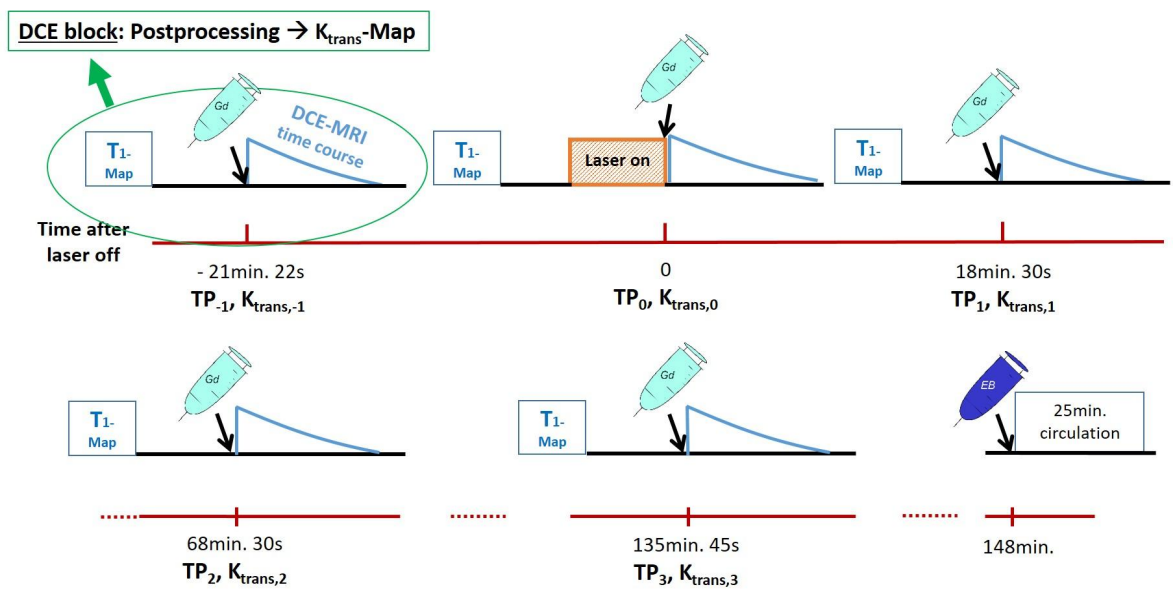

*Figure 2: Protocol of the multiple-bolus DCE experiment. A DCE-block (in green) consists of a T1 map and a standard DCE including CA injection. After CA injection, a DCE-MRI time course (in blue) is acquired from which a permeability map ($K_{trans}$) is obtained in a subsequent post-processing step.*

Rat in-vivo experiments were performed to acquire heating profiles for the subsequent DCE-experiments. After a short localizer acquisition, high-resolution anatomical images (FLASH, in-plane isotropic resolution of 136.7 µm) were acquired in order to localize the glass fiber end and define the region to be heated. These high-resolution images were ultimately used for superposition with the subsequently measured permeability maps.

In a necessary intermediate step, a $B_1$-map was acquired using the double angle method [45] so that flip angles could be adjusted in the ultimate post-processing part. Potential $B_1$-inhomogeneity effects due to the use of the surface coil could herewith be excluded.

The multiple-bolus DCE experiment consisted of several repetitions of a DCE-block. Each DCE-block comprised a $T_1$ map for calibration followed by a standard DCE measurement ($T_1$-weighted FLASH sequence). Here the signal dynamics in each voxel of $T_1$-weighted images are observed before, during and after a CA injection (0.1 ml/kg).

The map of the native relaxation time $T_{1,0}$ obtained from the initial $T_1$ maps together with the first pre-bolus image of the DCE measurement, enables the calculation of the baseline signal $S_0$ for each voxel using the well-known steady-state signal equation of a RF-spoiled FLASH sequence (see supplementary text S5). Knowing $S_0$, the CA effects on $T_1$ over time can be consequently calculated as a time dependent relaxation rate $r_1(t)$ by reformulating accordingly the same equation. This finally leads to the CA concentration dynamic in a voxel ($C_{vox}$) given by equation 2:

$$C_{Voxel}(t) = \frac{r_1(t) - 1/T_{1,0}}{r_{Gd}} \qquad \text{equation 2}$$

where $r_{Gd}$ is the relaxivity of the CA in $mM^{-1}s^{-1}$.

A Kermode-Tofts 2-compartment Pharmacokinetic model [46] is applied in the next step on the CA concentration time courses allowing to estimate the volume transfer coefficient $K_{trans}$ and the corresponding reflux rate $k_{ep}$ of the CA from the intravascular space to the extravascular extracellular space (EES).

The dynamic of the Gd concentration in the tissue $C_t$ can thus be described by equation 3:

$$\frac{dC_t(t)}{dt} = K_{trans} * C_p(t) - k_{ep} * C_t(t) \qquad \text{equation 3}$$

where $C_p$, the Gd concentration in blood plasma is obtained from the vascular input function (VIF) measured in the superior sagittal sinus (SSS) and subsequently corrected with the hematocrit factor as in [47].

The differential equation (equation 3) was analytically solved using a linear least squares function with nonnegativity constraints (MATLAB) for each image voxel, yielding finally to the $K_{trans}$ maps. $K_{trans}$ (in $min^{-1}$) is characterizing the capillary permeability of the Blood-Brain Barrier (BBB) by quantifying the Gd uptake in the tissue.

To track dynamic changes in BBB permeability, 5 DCE-blocks were performed as shown in figure 2.

The 5 timepoints when CA was injected in each block are denoted as $TP_{-1}$ to $TP_3$, whereby laser illumination occurred directly before $TP_0$. It is important to note that due to the dynamic nature of DCE-MRI acquisitions, the permeability maps should be interpreted as calculating one permeability value across the time course following each CA injection rather than at an instantaneous timepoint.

$T_1$ maps and DCE experiments were obtained using a RARE and a FLASH-sequence respectively. The geometry of all imaging experiments was 5 juxtaposed 1 mm- axial slices, matrix size = 61 x 85 and FOV = 25 mm x 35 mm resulting in an in-plane resolution of 410 µm x 412 µm. Also, relevant parameters for the DCE measurements were: TE = 1.3 ms, FA = 30 ° and TR = 30 ms, leading to a temporal resolution of 2.55 s per image.

# Results

*Temperature induced BBB opening*

To estimate the necessary thermal dosage (maximal temperature of brain tissue & duration of heat exposure) for BBB opening, cortical tissue was heated with increasing energy depositions during IR illumination. To stay within a clinically relevant parameter space, heating sessions did not exceed 10 minutes. Qualitative verification of BBB permeability increase was obtained by macroscopic EB extravasation at the site of IR-illumination. Figure 3A shows a representative image of the resulting discolorations and lack thereof corresponding to different thermal dosages. Based on the below described initial results, a thermal dosage of 43 ºC for 600 s was subsequently used during multi-bolus MR DCE measurements for precise determination of the threshold temperature for BBB opening.

*Different IR heating parameters for different thermal BBB opening thresholds*

To compare thermal dosages across experimental (IR laser) parameters, the total amount of deposited energy was plotted against the maximal tissue temperature in context of EB extravasation (see figure 3B). Total deposited energy was calculated by first determining the energy deposited during each pulse ($E_P$ (in Joule)) as follows:

$$E_P = P_P * PW \qquad \text{equation 5}$$

Where $P_P$ is the peak optical power (in Watt) of a pulse and $PW$ is the pulse width (in seconds). Subsequently, the total energy ($E$ (in Joule)) deposition was calculated by multiplying $E_P$ by the number of applied pulses. Thereby, determination of the total deposited energy ($E$) takes the different durations of exposure to the maximal temperature into account. Lowest energy values with simultaneous EB extravasation were determined at 1.07 J with a maximal tissue temperature of 44.5 ºC and an exposure time of 30 s.

The highest amount of deposited energy corresponding to no visible EB extravasation was 2.3 J spread out over 300s leading only to a maximal tissue temperature of 38.7 ºC. We conclude, that the leading factor for EB evasation is the tissue temperature achieved, but not the total amount of deposited energy, very much in agreement with the CEM43°C model (eqn. 7).

The average pulse power $P_{avg}$ (in Watt) of each IR heating session was calculated as follows:

$$P_{avg} = DC * P_P \qquad \text{equation 6}$$

where $DC$ is the duty cycle (in %). $P_{avg}$ was plotted against the maximal tissue temperature in the context of EB extravasation (see figure 3C). The maximal average pulse power at which no EB extravasation was observed, was at $P_{avg}$ = 23.4 mW with a maximal tissue temperature of 41.2 ºC at an exposure time of 60 s. The minimal average pulse power at which EB extravasation was observed was at 25.5 mW with a tissue temperature of 42.7 ºC for 600 s. A high correlation between maximal tissue temperature and $P_{avg}$ (R = 0.95), combined with the evidently strong role of a temperature threshold for BBB opening, makes $P_{avg}$ in the experimental set-up described here, a good predictor for EB extravasation.

However, any comparison of laser parameters with maximal tissue temperature is susceptible to inaccuracies due to the differences during in vivo settings such as the amount of liquid covering the illumination site, blood pressure and animal core temperature.

To evaluate the thermal tissue history only, the classical model for hyperthermic cell damage (cumulative equivalent minutes at 43 °C (CEM43°C)) [48] was plotted against the maximal tissue temperature during the respective heating sessions (figure 3D). CEM43°C calculates the duration of heating at 43 °C, equivalent to the damage caused by a given temperature using the following formula:

$$CEM43°C = \sum_{i=1}^{n} t_i \times R^{(43-T_i)} \qquad \text{equation 7}$$

where CEM43°C is the cumulative number of equivalent minutes at 43 °C, $t_i$ is the $i$-th time interval, $R$ is related to the temperature dependence of the rate of cell death ($R(T< 43 °C) = 0.25$, $R(T>43 °C) = 0.5$) and $T$ is the average temperature during time interval $t_i$. The lowest thermal dosage with EB extravasation observable was established by a maximal tissue temperature of 41.8 ºC for a duration of 300 s (CEM43ºC = 1.0). The highest thermal dose at which no EB extravasation of the targeted tissue was observed had a maximal tissue temperature of 41.9 °C with an exposure time of 60 s (CEM43ºC value of 0.2).

The close similarity of both maximal temperatures while observing different outcomes concerning EB extravasation underlines the important role of duration of heat exposure at this threshold temperature.

In summary, depending on the parameter used as the basis for analysis of EB extravasation, different thermal thresholds have been determined. The above described parameters point to either relatively long duration of heat exposure (≥300 s) with a tissue temperature of at least 41.8 ºC, or relatively high temperatures (≥44.5 ºC) with an exposure time of at least 30 s. We defined the thermal dosage during MR-DCE experiments conservatively as a combination of both (i.e. long exposure time (600 s) and high temperature (43 ºC)) analog to the predicted minimally necessary thermal dosage of EB extravasation based on $P_{avg}$.

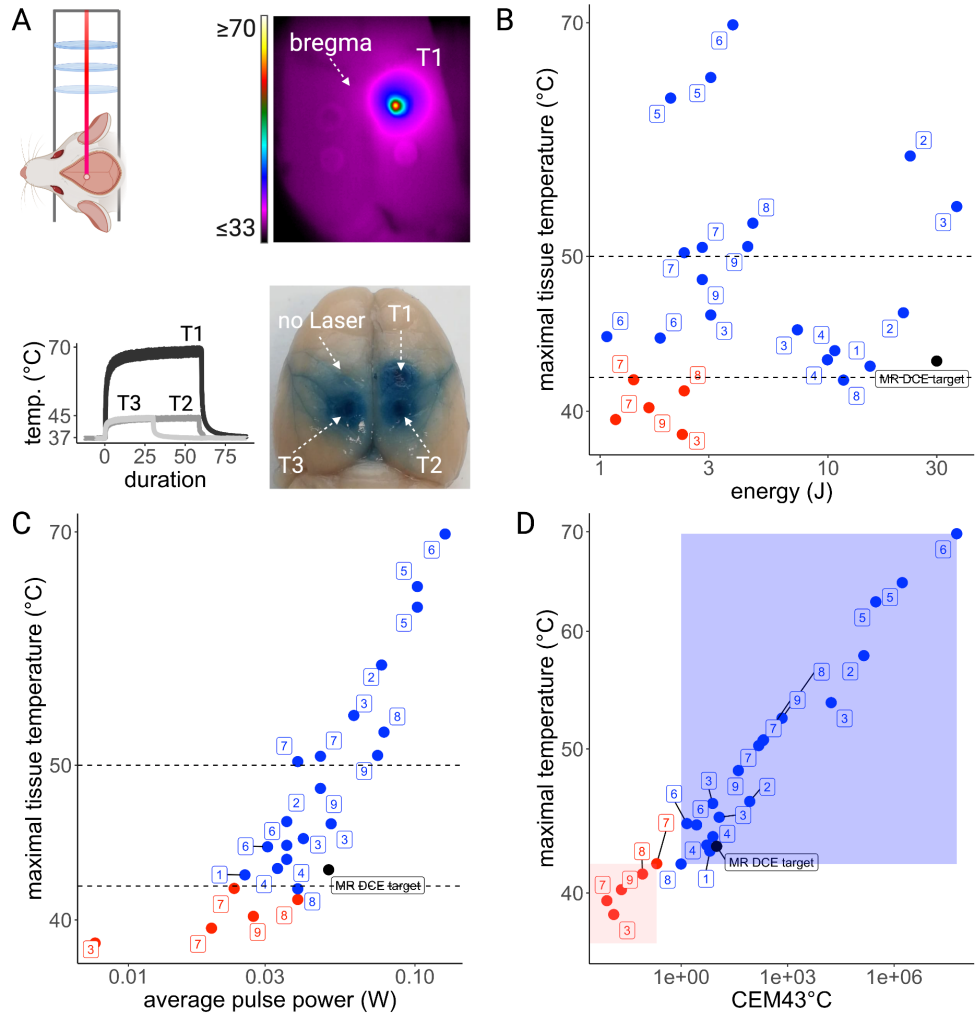

*Figure 3: IR induced Blood-Brain Barrier opening: IR laser projection through several optical lenses onto brain tissue was used to estimate the thermal dosage necessary for increased Blood-Brain Barrier permeability (A). Different thermal dosages were applied and measured via thermal camera. Permeability increase was observed by EB discoloration of the targeted tissue. The corresponding plots (B-D) show the results for positive (blue dots) and negative (red dots) EB extravasation in the context of different thermal dosage parameters. The numbers next to each sample point correspond to individual animal numbers. For reference, each plot includes the desired thermal dosage for subsequent BBB opening analysis based on MR-DCE measurements (black).*

## MR Thermometry: Calibration and Validation

The spatial and temporal temperature variations as a function of the supplied energy were investigated in the agarose phantom and in post-mortem and in-vivo condition of the rat. Especially the temperature behavior at the glass fiber end and the heat propagation in space were of particular interest.

The peak temperature in the ROI (250 µm x 250 µm) at the glass fiber end showed as expected a linear dependency on increasing laser power: ΔT is plotted versus increasing DC in figure 4. A linear regression applied to the course (figure 4, left) shows a temperature increase of 2.46 K +- 0.46 K per 1% DC ($P_{av}$ = 5 mW) in the agarose phantom. A similar behavior is observed with the thermal camera experiment with a 2.26 K +- 1.2 K increase per 1% DC. Both the MR thermometry and the thermal camera measurements delivered similar values (8.1% deviation) Similarly, post-mortem and in-vivo measurements showed a linear behavior with a temperature increase of 2.61 K +- 0.6 K and 2.14 K +- 0.22 K per 1% DC respectively (figure 4 right). A difference of 18% was observed between post-mortem and in-vivo measurements.

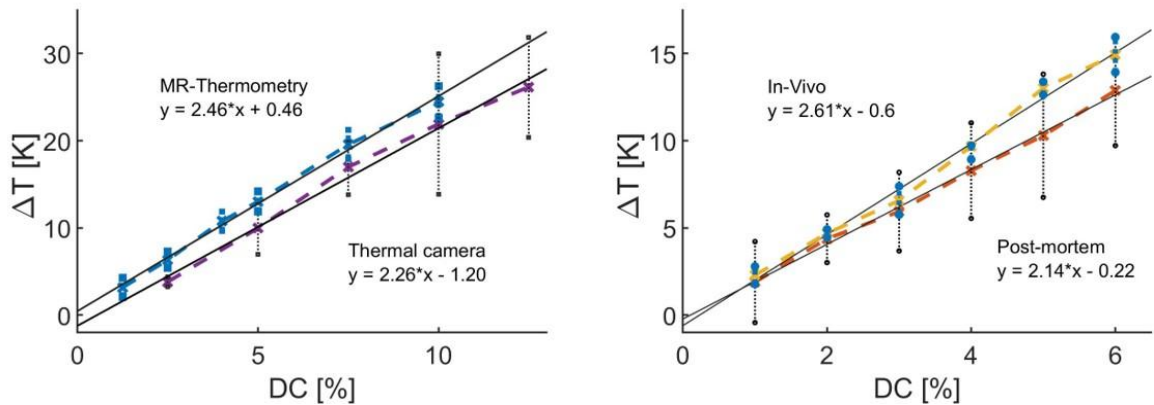

*Figure 4: The temperature increases detected in the voxel following the glass fiber end are plotted against the increasing DC in %. MR-Thermometry and Thermal Camera phantom results are compared (left) as well as the results for the rat in post-mortem and in-vivo conditions (right). A 1% DC step is equivalent to a $P_{av}$ increase of 5 mW.*

The heat dissipation was investigated in 4 dimensions. First, the 2D plane of the slice in which the glass fiber termination was located, was analyzed. Results for the ΔT decrease over distance in the agarose phantom as well as isolines for chosen representative temperatures (3 K, 4.5 K, 6 K and 9 K) are shown for DCs of 5% ($P_{av}$ = 25 mW) and 10% ($P_{av}$ = 50 mW) in figure 5a and 5c respectively. In order to get a clearer representation of the heat dissipation over distance in the plane, a 2D representation of the isolines is also shown in figure 5b and 5d to help. Exponential functions were fitted to the data in all three directions shown in figure 5a and 5c: 13.17*exp(-1.28*x)+0.53 (+x), 13*exp(-1.21*x)+0.70 (-y) and 13.41*exp(-1.17*x)+0.69 (+y) with R² = 0.985, 0.986 and 0.986 respectively. A fit to the mean value of the experimental data was used in the latter part to locate the 6K-isoline (equivalent to a temperature of 43 °C) in the permeability maps: 13.19*exp(-1.22*x)+0.64 ( R² = 0.991).

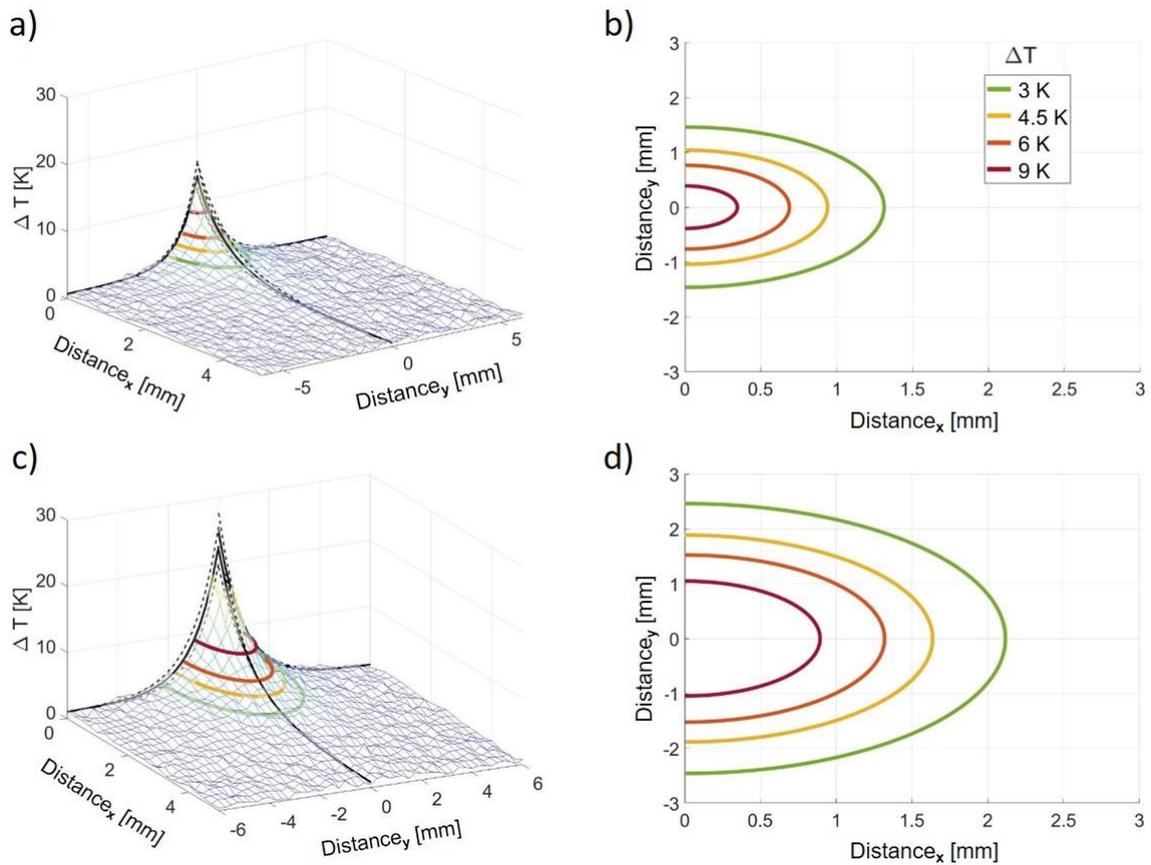

*Figure 5: The ΔT behavior in the image slice where the glass fiber terminates is shown on the left for a DC of 5% and 10% respectively. The 3K-, 4.5K-, 6K- and 9K- isolines are also shown while the glass fiber end is situated at (0,0,0). A 2D representation of the very same isolines is shown right, facilitating the readability.*

The heat dissipation across image slices is shown in figure 6. The heat map of the slice next to the glass fiber terminal is displayed together with the 3K-, 6K- and 9K-isolines for subsequent slices. The typical exponential course can be observed as well in the slice direction.

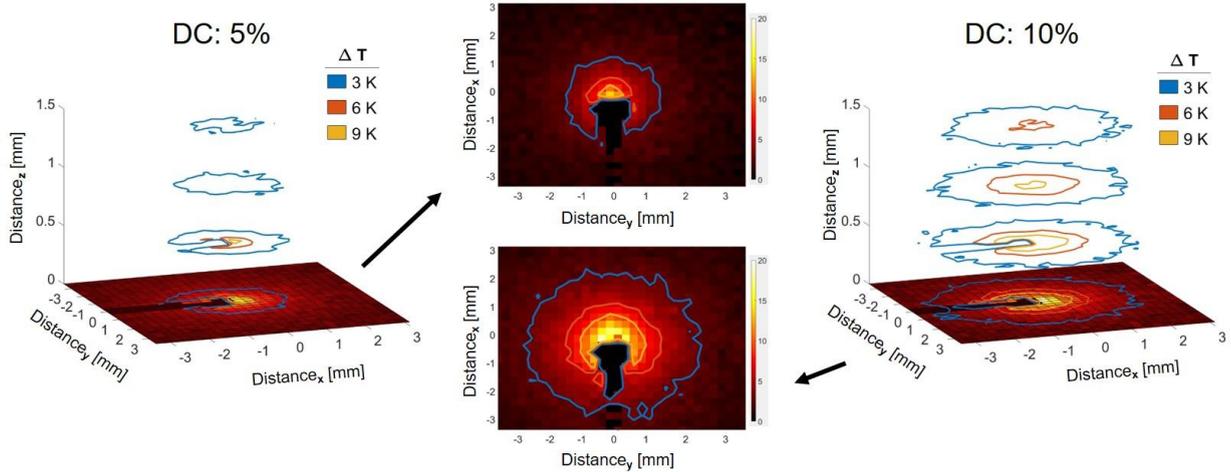

*Figure 6: Heat maps are shown for the slice in front of the glass fiber termination for a DC of 5% ($P_{av}$ = 25 mW) and 10% ($P_{av}$ = 50 mW) respectively. Note that the black region corresponds to artifacts from the glass fiber. The 3K-, 6K- and 9K- isolines are shown for the very same slice and the 3 next slices 0.5 mm, 1 mm and 1.5 mm away.*

In addition to the 2D and 3D measurements, changes of ΔT over time were examined (figure 7). In particular, the dynamic temperature changes for the voxels near the laser output were investigated in the post-mortem and in-vivo conditions of the rat (Distances: 0 mm, 0.5 mm, 1 mm and 2 mm). As the focus here was on the signal behavior over time, a higher temporal resolution of 5.12 s per phase maps was chosen, leading to a spatial resolution of 500 µm. DC values from 1% to 6% in 1% steps corresponding to a $P_{av}$ of 5 mW to 30 mW with 5 mW steps respectively were considered. A faster decay of the ΔT-plateaus over distance as well as flatter plateaus were observed in in-vivo conditions in comparison to the post-mortem experiments.

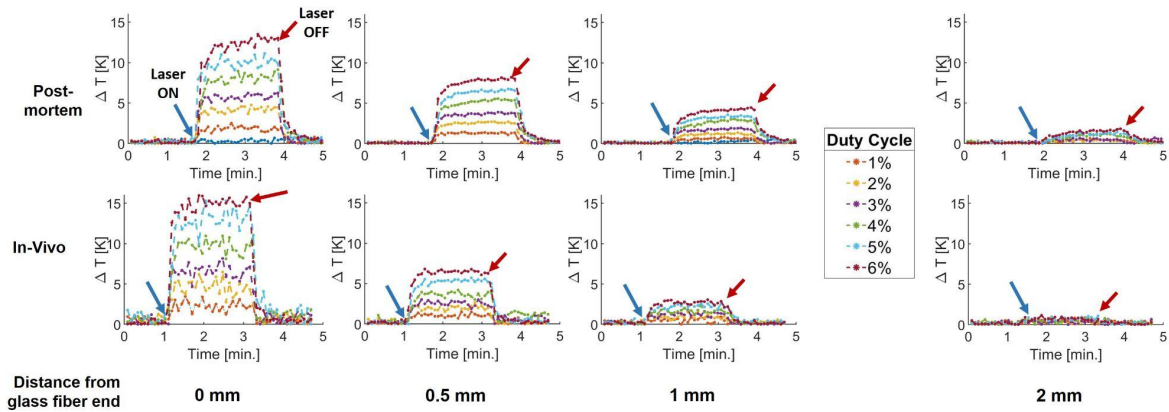

*Figure 7: The ΔT time courses are plotted at the terminal of the glass fiber as well as at 0.5 mm, 1 mm and 2 mm in front of it. The experiments were repeated for several DC between 1% and 6% for post-mortem (top line) and in-vivo (bottom line) conditions of the rat. Blue and red arrows show where the laser is turned on and off respectively.*

Spatial thermal dissipation can be illustrated by superimposing the thermal heat maps superimposed on high-resolution anatomical MR images. In addition, temperature isolines show the area where an opening of the BBB and therefore a higher permeability is expected.

Figure 8 shows the measured heat map for DC 6% ($P_{av}$ = 30 mW) in the in-vivo condition of the rat for the slice in front of the glass fiber terminal, along with a mask showing temperatures above 1 K and 3 K-, 6K- and 9K- isolines. As the rat body temperature was maintained at 37 °C, the 6 K- isoline is of particular interest as it exceeds the identified lower limit necessary to open the BBB. As expected, no temperature increase is detected at the site where the sham glass fiber is situated. As a guide structure was used for the animal experiments, the glass fiber enters the brain with a predefined angle of 20° leading to the in-plane deviation observed in figure 8.

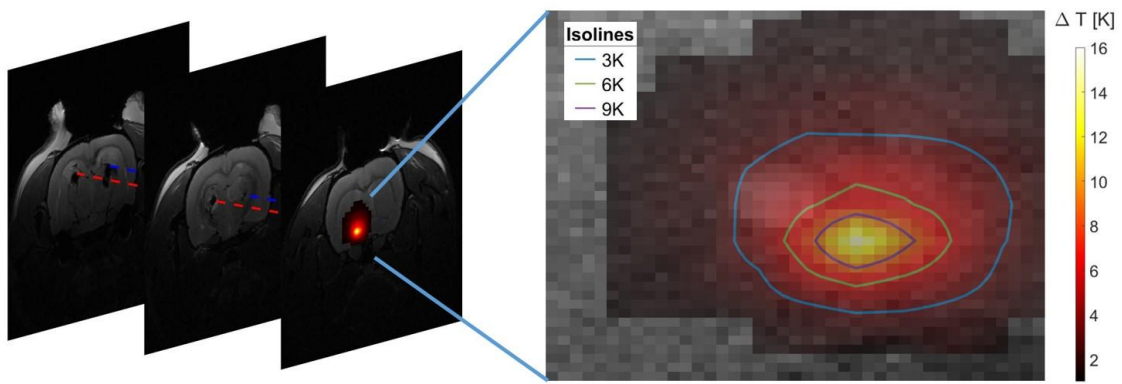

*Figure 8: Three slices of the anatomical images are shown on the left. The red and blue dotted line shows the angle of the laser and sham glass fibers respectively, both ending in the third slice where the heat map is also shown. On the right, the heat map together with the 3K- (blue), 6K- (green) and 9K- (purple) isolines superimposed on the anatomy image of the rat are shown.*

Finally, a closer look at the evolution of the 6K-isoline with increasing DC was investigated (figure 9). A DC of at least 3% ($P_{av}$ = 15 mW) according to our assumption is needed to open the BBB. Areas of 0.02, 0.36, 0.99 and 1.65 mm² for DC values from 3% to 6% ($P_{av}$ 15 mW to 30 mW) are expected for the latter permeability maps.

Even if much effort is put into maintaining the body temperature of the rat constant at 37 °C, there may be variations across individuals as well as dependencies on the animal's plane of anesthesia or stress level. Therefore, we also investigated the 5K- and 7K- isolines corresponding to the ΔT needed for core temperatures of 38 °C and 36 °C respectively. This predicts, for example with a DC of 6%, an area of increased permeability of 2.71 and 1.04 mm² for a ΔT of 5 K and 7 K, respectively.

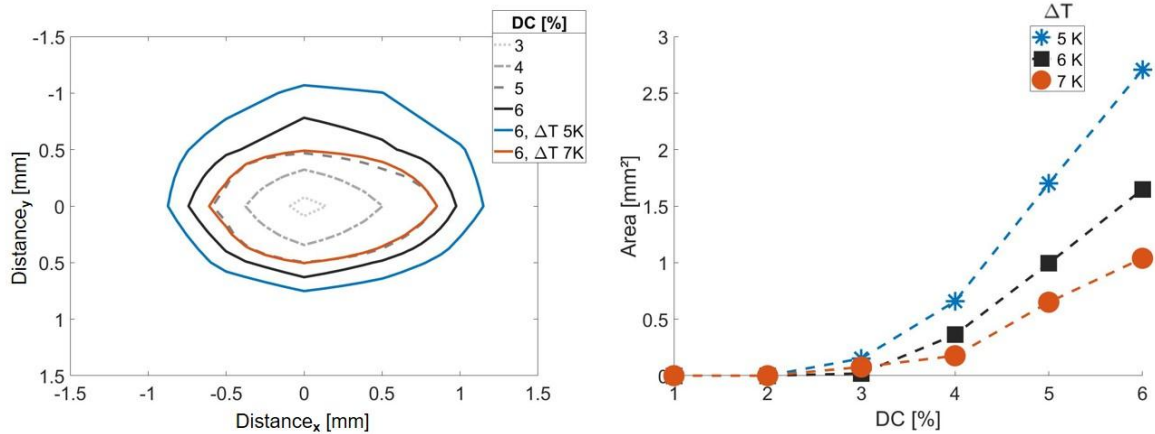

*Figure 9: The 6K-isolines for DC going from 3% ($P_{av}$ = 15 mW) to 6% ($P_{av}$ = 30 mW) in 1% steps ($P_{av}$ = 5 mW) are shown on the left. Additionally, the 5K- and 7K- isolines for a 6% DC is shown. On the right the predicted area of increased permeability for DC from 1% to 6% is shown for ΔT = 5 K, 6 K and 7 K, required for rat body temperatures of 38 °C, 37 °C and 36 °C respectively.*

*Gd-induced signal behavior in a multiple-bolus DCE experiment*

While analyzing the signal evolution over time in the multiple-bolus DCE experiment, we differentiated 4 typical signal behaviors. Thus, on the one hand, a voxel situated at a vein or artery location shows a typical vascular input function (VIF) shape (yellow in figure 10): Strong signal increase is followed by a steep decay as the CA is flowing through the voxel. Voxels containing both blood flow and static tissue show a slower descending slope after the bolus occurs but the signal still goes back to its initial value after some time (purple in figure 10).

On the other hand, the signal behavior in voxels placed e.g. in the white matter typically shows a noisy signal (red in figure 10) where the bolus passing through the capillaries is barely seen. Here, the mean signal value remains unchanged over time. If the BBB is open, an accumulation of CA and with it an augmentation of the signal over time will occur ($S_{-1} \to S_3$, blue in figure 10). Typically the signal after injection remains quickly reaches a plateau.

Those different signal time courses will be used while analyzing the $K_{trans}$-maps in the following to mask voxels of no interest or rather to characterize those of interest.

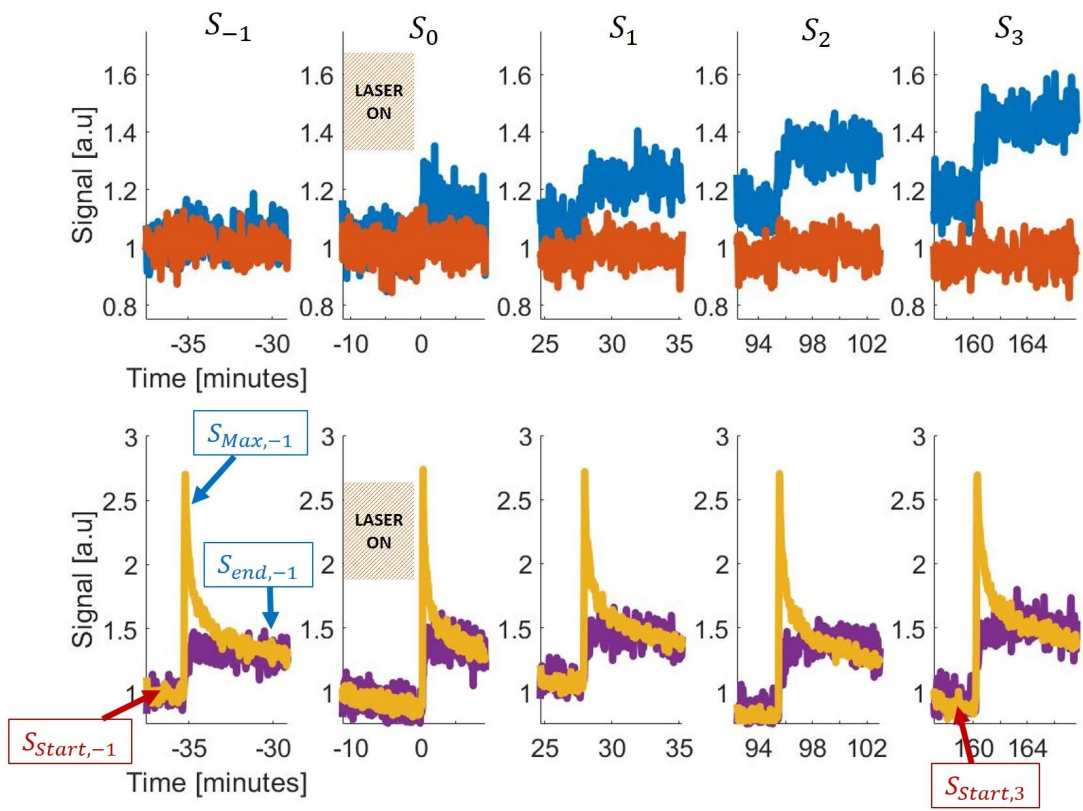

*Figure 10: Top row shows the temporal signal evolution in a tissue voxel with (blue) and without (red) permeability enhancement. The same is shown on the bottom line for a voxel completely (yellow) or partly (purple) situated in a larger vessel.*

*BBB permeability induced by the infrared laser made visible in $K_{trans}$-maps*

A Kermode-Toft fit as described in the methods part is applied to the acquired DCE signals in order to calculate $K_{trans}$-maps. With each bolus, a new map is calculated and thus the temporal evolution of the permeability in tissue can be monitored.

For the mild hyperthermia experiment, a permeability increase over time is observed at the location where energy is deposited (red circle in figure 11) while at the location of the sham GF (blue circle), no increase is observed. Also in the surrounding slices no Gd accumulation could be detected. This shows that the permeability increase is achieved locally through the laser illumination and not by some bleeding effects that might occur during the animal surgery.

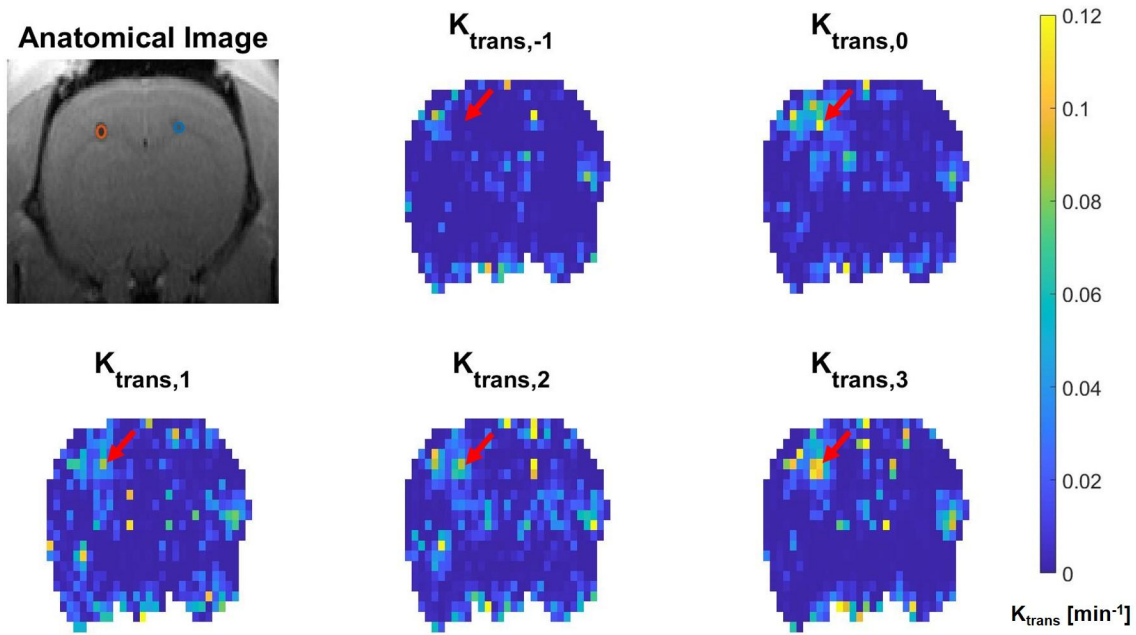

*Figure 11: Anatomical image of the implanted rat's brain: The working ends of the laser-connected glass fiber and the sham are shown in red and blue, respectively. Subsequently, the raw $K_{trans}$ maps are shown for all timepoints $TP_{-1}$ until $TP_3$. Red arrows show where a permeability increase is expected.*

A systematic use of several masks based on the signal time courses (figure 12) was applied in order to verify the specificity of the permeability increase effects through the temperature augmentation.

In the first step, every voxel showing no absolute permeability after the illumination ends ($TP_0$) was removed. As $K_{trans}$-maps show only an absolute value, $\Delta K_{trans}$ reflecting the relative permeability increase was introduced. $\Delta K_{trans}$ is defined as the difference between the $K_{trans}$-maps at any TP and $K_{trans,-1}$, describing the brain permeability before the laser is turned on ($TP_{-1}$). The first mask applied ($Mask_1$) removed all voxels showing a nulled $\Delta K_{trans}$ at any TP (first line in figure 12).

As can be seen e.g. in figure 12d ($\Delta K_{trans,0}$), there are several spatially distributed voxels showing a permeability increase. In addition to those induced through the mild hyperthermia, also those containing larger vessels show intrinsically a higher $\Delta K_{trans}$ value. Here, one might expect $\Delta K_{trans}$ shall be zero. However, even if the boli were similar, some differences in quantity and injection velocity will still lead to the detection of a different increased permeability at any TP.

Therefore, a second mask (Mask$_2$) was created, using the specificity of the temporal evolution of the vascular signal (see figure 12). The difference between the maximum of the undisturbed original signal $S_{-1}$ over time and the mean value of the last 50 measurement points from $S_{-1}$ was calculated as in equation 8.

$$F_1 = \frac{S_{-1,max} - S_{-1,end}}{S_{-1,max}} \qquad \text{equation 8}$$

All voxels showing a signal decrease higher than the mean value of $F_1$ were masked (Figure 12c, $F_1$).

In the last step, we investigated for the remaining voxels if the initial DCE signal preceding the first bolus injection is increasing over time as it would while Gd gets stored. Thus, a mask (Mask$_3$) based on the difference between the signal mean value of the first 40 measurement baseline points at TP$_{-1}$ ($S_{-1}$) and TP$_3$ ($S_3$) was used as in $F_2$:

$$F_2 = \frac{S_{-1,start} - S_{3,start}}{S_{-1,start}} \qquad \text{equation 9}$$

Finally, as can be seen in figure 12, only voxels around the location of mild hyperthermia remain.

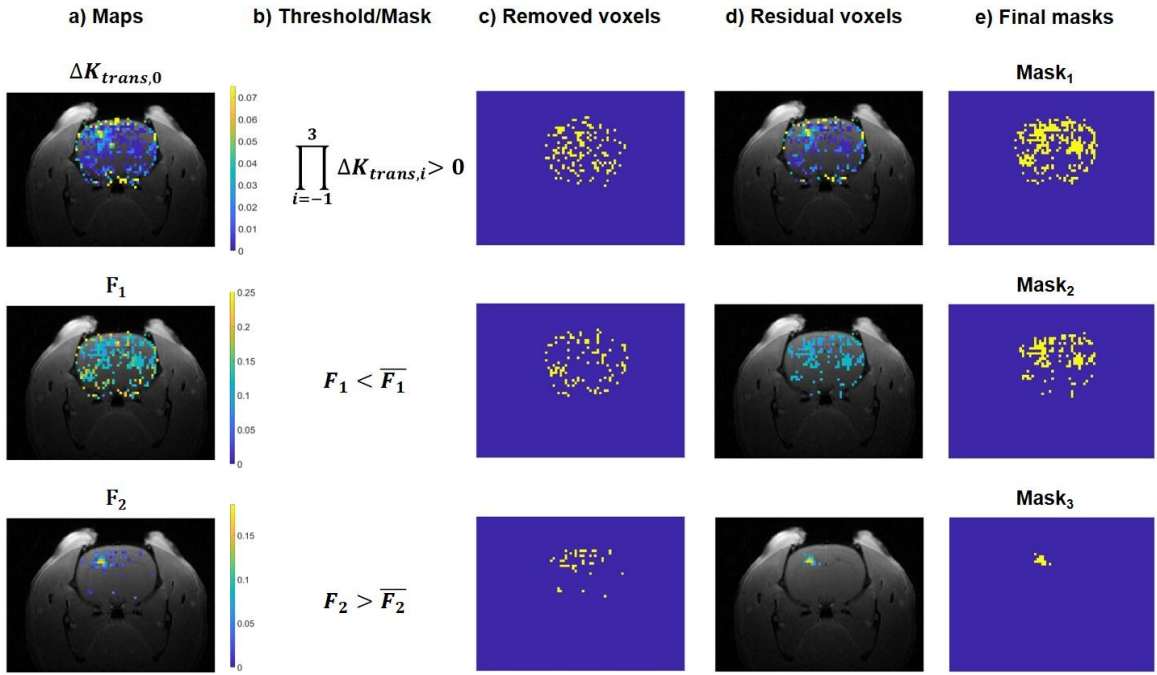

*Figure 12: Several masks were used to prove the specificity of the method. Voxels not showing an increased permeability at all TPs (ΔK$_{trans,0}$) as well as those showing an increased ΔK$_{trans}$ value due to larger vessels (F$_1$) are masked. A final mask masked filtrated voxels not showing increasing signal between TP$_0$ and TP$_3$ (F$_2$).*

After we could show that the BBB opening occurs through local mild hyperthermia using laser illumination, we investigated the limits of the temperature needed for the BBB opening . As the rat showed a constant body temperature of 37 °C, we overlaid the 6K-isoline on the ΔK$_{trans,3}$ map. Here, the isoline was calculated from the exponential fit of the mean in-vivo thermometry results for the corresponding 6% DC as introduced before. Also, a gaussian function was fitted on the permeability map as a normal distribution can be reasonably assumed. A strong correlation between the expected hyperthermia and the BBB-area opening could be observed as is displayed in figure 13.

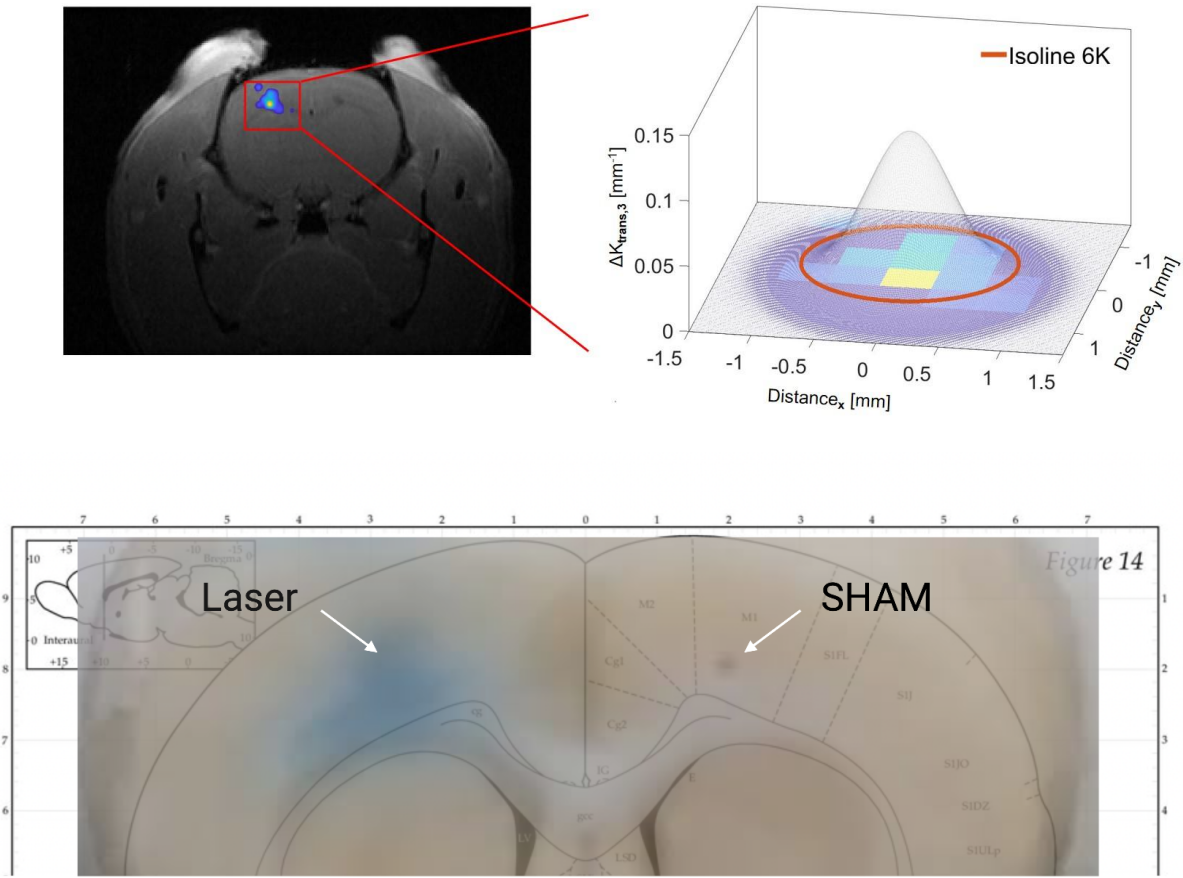

*Figure 13: ΔK$_{trans,3}$ was superposed to a high-resolution anatomical image of the rat brain. A zoom in of a gaussian fit over ΔK$_{trans,3}$ is shown right together with the 6K-isoline, expected spatial limit of the BBB opening (top row). Overlay of brain slice with anatomical location (bottom row). Post-mortem EB extravasation can be observed at the site of IR illumination (left) but not at the location of the SHAM fiber (right).*

# Discussion

Several approaches to transiently increase BBB permeability in order to locally enhance drug concentrations have been investigated over the last decades. Tissue damages and off-target side effects remain the limiting factor for these methods [28][29]. Therefore, extensive research has been dedicated to the exploration of alternatives minimizing the issues associated with existing techniques.

In this study, we used highly localized IR illumination to induce BBB opening. Although implantation of glass fibers to deliver heat to the targeted brain tissue cannot be considered non-invasive, associated damages can be minimized as has been shown for example for electrode implantation into the brain [49–51]. The induction of BBB opening itself however, is entirely based on the transient and mild increase of local tissue temperature. The 1470 nm wavelength was selected in order to maintain as much as possible control over the targeted area of interest and reduce the heating impact on the immediate surroundings. Lower wavelengths lead to less localized heating and thereby to more off-target tissue effects (see supplementary figure S3 for a representative comparison). Also the absorption coefficient values at 1470 nm are a few orders higher for water than for hemoglobin [52].

To minimize tissue damage and limit the duration of BBB permeability increase, it is crucial to determine the necessary thermal dosage to induce BBB opening. To investigate that, we have conducted several experiments on the relation of thermal dosage and BBB opening based on extravasation of EB at targeted tissue sites. These initial experiments enabled us to further home in on the heat induced BBB threshold dosage. In general, longer heating durations of several minutes (at least 5) in combination with relatively low tissue temperatures (≥ 41.8 ºC) and conversely, low durations (≥ 30 s) of high temperatures (≥ 44.5 ºC) are necessary to effectively induce BBB opening. The thermal threshold for any of the analyzed parameters (i.e. energy and average pulse power) seems to corroborate the CEM model with a critical tissue temperature of 43 ºC [48]. If it is assumed that (transient) thermal tissue damage is necessary for effective BBB opening (disruption), this temperature is in accordance with previously described thermal thresholds for cell damage [53–55]. However, for precise threshold determination of BBB disruption, macroscopic EB extravasation becomes cumbersome. EB extravasation is dependent on several factors such as blood pressure and individual vessel density and is generally not suited for quantitative analysis related to the BBB [56]. Further, since EB extravasation analysis can only be done post-mortem, no conclusions can be drawn to the temporal progression of BBB opening.

MR thermometry was used to retrace tissue temperature during subsequent BBB permeability investigations using multiple bolus DCE measurements. Phantom MR thermometry measurements for IR illumination with increasing DCs have been compared and verified with thermal camera measured data using the same laser parameters.

Relatively large error bars have been observed for thermal camera based IR laser calibration. We associate this observation with slight inconsistencies of the heated bare agarose surface used in the experimental setup (see supplementary figure S3).

Further, slight deviations for the MR thermometry results between post-mortem and in vivo measurements have been observed. Because these differences increase with increasing DC (i.e. maximal temperature), we associate this effect with circulation induced cooling at the site of heating in-vivo.

The obtained results from EB extravasation experiments have been used as the basis for subsequent MR-DCE measurements. The accuracy of the laser heating in phantom (agarose), post-mortem and in-vivo measurements were systematically analyzed, eventually enabling a calibration and therefore precise control of the highly-localized heating area. The desired thermal dosage of 43 ºC for 10 minutes was subsequently achieved by applying IR illumination with 6% DC ($P_{av}$ = 30 mW) and a peak pulse power ($P_p$) of 0.5 W assuming a regular rat core temperature of 37 °C and a similar brain tissue temperature in-vivo. According to in-vivo MR thermometry results, this temperature is achieved in an area of 1.65 mm² around the glass fiber terminal. Correspondingly, we assumed increased BBB permeability within this area.

Through the use of multiple-bolus DCE, we were able to verify this assumption by overlaying the high-resolution anatomical image, with the area covered by the MR thermometry predicted 6K-isoline and the $\Delta K_{trans}$-map which characterizes the capillary permeability of the BBB.

Due to the terminal nature of the in-vivo MR DCE measurements, BBB permeability was not investigated beyond 2h post IR illumination. In the literature, different durations for post interventional BBB opening have been presented ranging from minutes[57] to several hours[58] which seems to be dependent on the method to induce permeability increase and the type of the investigated permeability marker [22]. The duration of heat induced BBB opening is highly dependent on the applied thermal dosage. During thermal tissue ablation using high temperatures an increased BBB permeability has been reported for up to several weeks[59,60]. For the here described mild thermal dosage, we observed increased permeability for at least 2h. Future experiments should validate the reversibility of the method.

# Conclusion

In this work, we investigated the feasibility of a highly localized increase of the BBB permeability by mild hyperthermia using IR illumination which is a promising approach as it may alleviate side-effects associated with established and novel methods [61]. A crucial information to improve heat induced BBB opening, is the determination of the minimally necessary tissue temperature which was addressed in this study. Synchronizing the results obtained from MR thermometry and multiple bolus DCE during in-vivo IR illumination, we determined that a cortical temperature increase of 6 K suffices for effective BBB opening assuming a core temperature of 37 ºC.

# Supplementary

**Supplementary Table 1: Parameters and results for thermal blood brain barrier opening threshold estimation**

| animal No | peak pulse power $P_P$ (W) | average pulse power $P_M$ (W) | pulse width PW (s) | pulse energy $E_P$ (J) | duty cycle (%) | frequency (Hz) | no. of pulses | duration (s) | energy E (J) | max. temperature (ºC) | Evans' Blue extravasation | ROI |
|---|---|---|---|---|---|---|---|---|---|---|---|---|
| 1 | 10,2 | 0,0255 | 0,00005 | 0,00051 | 0,25 | 50 | 30000 | 600 | 15,3 | 42,7 | Yes | 3 |
| 2 | 10,2 | 0,0765 | 0,00015 | 0,00153 | 0,75 | 50 | 15000 | 300 | 22,95 | 57,8 | Yes | 1 |
| 2 | 10,2 | 0,0357 | 0,00007 | 0,000714 | 0,35 | 50 | 30000 | 600 | 21,42 | 46,1 | Yes | 2 |
| 3 | 10,2 | 0,00765 | 0,000015 | 0,000153 | 0,075 | 50 | 15000 | 300 | 2,295 | 38,7 | No | 1 |
| 3 | 10,2 | 0,0612 | 0,00012 | 0,001224 | 0,6 | 50 | 30000 | 600 | 36,72 | 53,7 | Yes | 2 |
| 3 | 10,2 | 0,051 | 0,0001 | 0,00102 | 0,5 | 50 | 3000 | 60 | 3,06 | 45,9 | Yes | 3 |
| 3 | 10,2 | 0,0408 | 0,00008 | 0,000816 | 0,4 | 50 | 9000 | 180 | 7,344 | 45 | Yes | 4 |
| 4 | 10,2 | 0,03315 | 0,000065 | 0,000663 | 0,325 | 50 | 15000 | 300 | 9,945 | 43,1 | Yes | 2 |
| 4 | 10,2 | 0,0357 | 0,00007 | 0,000714 | 0,35 | 50 | 15000 | 300 | 10,71 | 43,6 | Yes | 3 |
| 5 | 10,2 | 0,102 | 0,0002 | 0,00204 | 1 | 50 | 1000 | 20 | 2,04 | 62,8 | Yes | 2 |
| 5 | 10,2 | 0,102 | 0,0002 | 0,00204 | 1 | 50 | 1500 | 30 | 3,06 | 64,7 | Yes | 3 |
| 6 | 10,2 | 0,1275 | 0,00025 | 0,00255 | 1,25 | 50 | 1500 | 30 | 3,825 | 69,8 | Yes | 1 |
| 6 | 10,2 | 0,0357 | 0,00007 | 0,000714 | 0,35 | 50 | 1500 | 30 | 1,071 | 44,5 | Yes | 2 |
| 6 | 10,2 | 0,0306 | 0,00006 | 0,000612 | 0,3 | 50 | 3000 | 60 | 1,836 | 44,4 | Yes | 3 |
| 7 | 7,8 | 0,0234 | 0,00006 | 0,000468 | 0,3 | 50 | 3000 | 60 | 1,404 | 41,9 | No | 1 |
| 7 | 7,8 | 0,0585 | 0,00015 | 0,00117 | 0,75 | 50 | 3000 | 60 | 3,51 | 50,3 | Yes | 2 |
| 7 | 7,8 | 0,0468 | 0,00012 | 0,000936 | 0,6 | 50 | 3000 | 60 | 2,808 | 50,6 | Yes | 3 |
| 7 | 7,8 | 0,0195 | 0,00005 | 0,00039 | 0,25 | 50 | 3000 | 60 | 1,17 | 39,5 | No | 4 |
| 8 | 7,8 | 0,0234 | 0,00006 | 0,000468 | 0,3 | 50 | 3000 | 60 | 1,404 | 41,2 | No | 2 |
| 8 | 7,8 | 0,078 | 0,0002 | 0,00156 | 1 | 50 | 3000 | 60 | 4,68 | 52,4 | Yes | 3 |
| 8 | 7,8 | 0,0273 | 0,00007 | 0,000546 | 0,35 | 50 | 15000 | 300 | 8,19 | 41,8 | Yes | 4 |
| 9 | 7,8 | 0,0741 | 0,00019 | 0,001482 | 0,95 | 50 | 3000 | 60 | 4,446 | 50,7 | Yes | 1 |

| 9 | 7,8 | 0,0195 | 0,00005 | 0,00039 | 0,25 | 50 | 3000 | 60 | 1,17 | 40,2 | No | 2 |

*Table S1: Summary of the IR laser parameters and thermal camera measured maximal tissue temperature for each animal and respective ROI used for EB extravasation experiments.*

## Supplementary figure S1: Thermal threshold estimation of Blood-Brain Barrier opening based on Evans' Blue extravasation

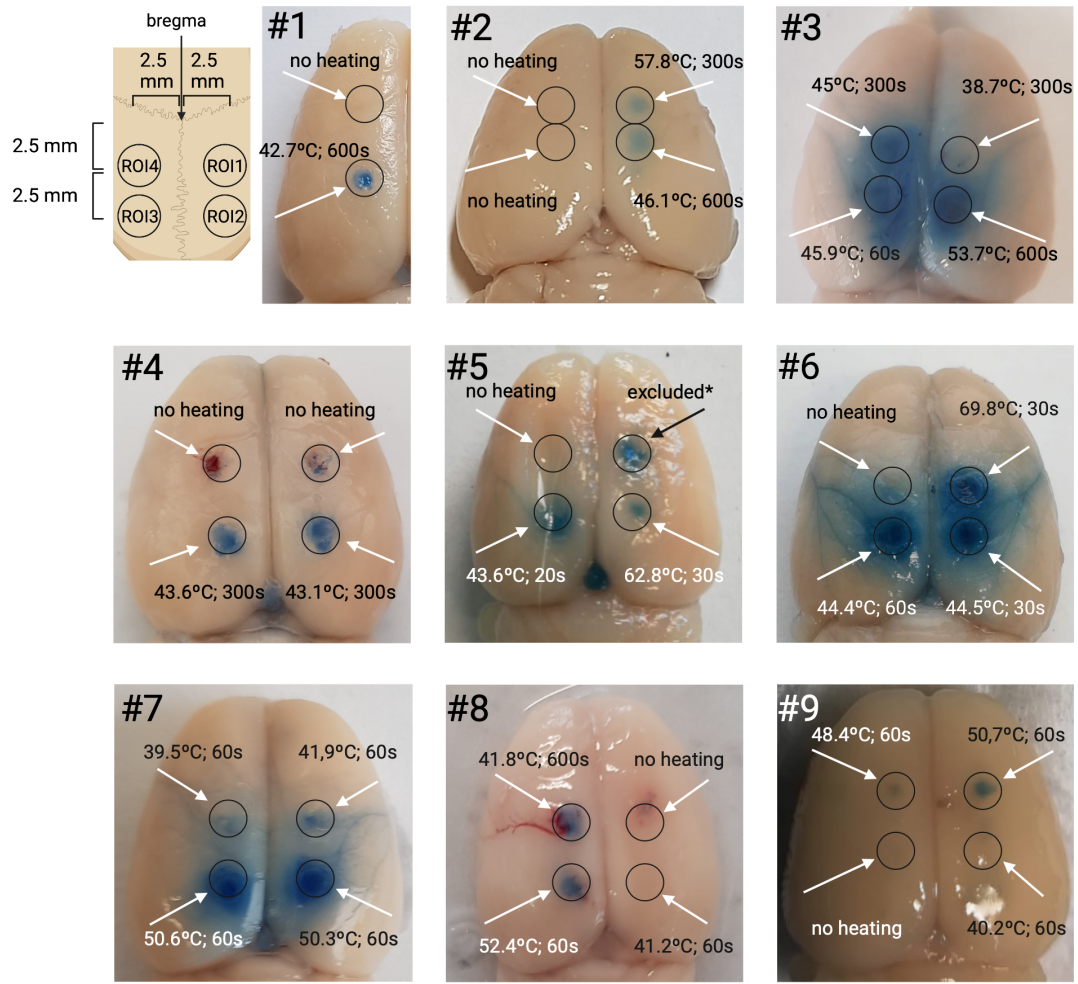

Supplementary figure S1: Evans' Blue extravasation as response to different thermal dosages. EB extravasation was determined by unambiguous discoloration of the targeted tissue. For each animal, 4 burr holes were drilled at the same location of the skull. No heating (no laser) or low thermal dosages and consequently no discoloration at the respective sites served as control. ROI1 (upper right) of animal #5 has been excluded from analysis because thermal camera data was not obtained.

## Supplementary figure S2: Experimental set up for IR laser calibration using a thermal camera

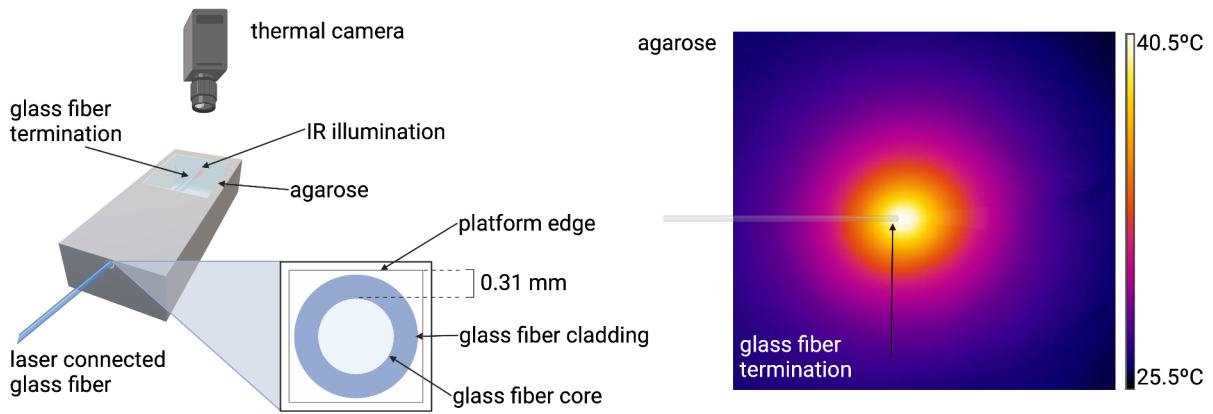

Supplementary figure S2: Thermal camera measured temperature calibration of IR illumination: A laser connected glass fiber was guided into a pool of agarose (0.6%) with a defined distance of the edge of the glass fiber termination to the agarose surface (left). For simplicity, only one distance of the agarose surface to the horizontal edge of the glass fiber is shown. For calibration however, a total of 7 distances corresponding to 0.31, 0.81 1.47, 1.97, 2.63, 3.13, 4.29 mm were used. The agarose temperature profile was measured at different duty cycles by a thermal camera positioned perpendicular to the agarose surface. An exemplary screenshot of the thermal camera measured temperature during IR heating (duty cycle = 10%) at 180 s of heating is shown on the right.

***Supplementary figure S3: MR generated heating maps for two different wavelengths: λ = 1470 nm and λ = 980 nm***

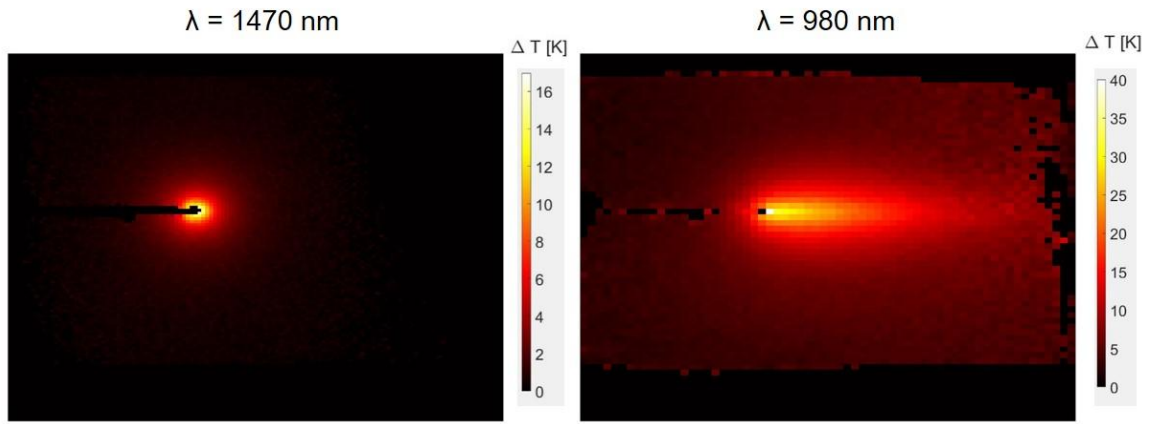

*Supplementary figure S3: Representative MR-thermometry heating maps for λ = 1470 nm (DC: 10%, $P_{peak}$ = 500 mW, τ = 1 μs, laser on 10 min.14s, $E_{tot}$ = 30.7 J) and 980 nm (DC: 75%, $P_{peak}$ = 500 mW, τ = 1 μs, laser on 2min.33s, $E_{tot}$ = 57.375 J)) are shown. The heating for 1470 nm is highly localized, almost referring to a heating tip. 980 nm in contrast shows a much higher spatial heat distribution and therefore impact on tissue, almost resulting in a heating torch.*

*Supplementary text S1: DCE post-processing*

The steady-state signal of an RF-spoiled FLASH sequence without influence of the CA is given by equation S1

$$S(t) = S_0 * \frac{\sin\theta * (1 - E_1)}{(1 - \cos\theta * E_1)} * E_2^*$$

where

$$E_1 = e^{-TR/T_1} \text{ and } E_2^* = e^{-TE/T_2^*}$$

In strong $T_1$ weighted images which is the case here, $E_2^*$ can be considered ≈ 1.

Taking the CA effects on $T_1$ over time into account go back to the fact that $T_1$ becomes $T_1(t)$ in equation S1. The time dependent relaxation rate $r_1(t)$ needed to obtain the voxel concentration (see Equation 2) can therefore be calculated by solving equation S1 for $1/T_1$:

$$r_1(t) = - \frac{\ln \frac{\left(\frac{S(t)}{S_0} - \sin\theta\right)}{\left(\frac{S(t)}{S_0} * \cos\theta - \sin\theta\right)}}{TR}$$

**Acknowledgment and Funding**

This work was funded by the German Federal Ministry of Education and Research, as project FMT - Functional Magnetotherapy, contract number 13GW0230A and in part by the project NEUROPHOS - 13GW0155C. The authors thank T. Klotzbücher and P. Detemple of Fraunhofer IMM for expert optical assistance.

The authors thank the Core Facility AMIR$^{CF}$ (DFG-Resources N° RI_00052) for support in MRI imaging.

SB, OB, CM, PS and UGH conducted and conceived experiments. UGH and DvE supervised the study. UGH directed the project FMT. UGH and PL secured funding for this project. All authors contributed to discussing, writing and revising the manuscript and approved the final version.